\documentstyle[aps,preprint,epsf]{revtex}
\begin{document}

\draft

\title{Kinetic Theory of dilute gases under  nonequilibrium conditions}

\author{ Hisao Hayakawa\footnote{hisao@yuragi.jinkan.kyoto-u.ac.jp},
 Taka H.  Nishino}
\address{Department of Physics, Yoshida-South Campus, Kyoto University, Kyoto 606-8501}
\author{and Kim Heyon-Deuk}
\address{Department of Chemistry, Kyoto University, Kyoto 606-8502}

\date{\today}

\maketitle

\begin{abstract} 
The significance of the recent finding of
the velocity distribution function of the steady-state Boltzmann equation under a steady heat current obtained by
 Kim and Haykawa (J. Phys. Soc. Jpn. {\bf 72}, 1904 (2003)) is discussed.   
Through the stability analysis, it seems that the steady solution is
 stable. One of possible
applications to the nonequiliburium Knudsen effect in which one cell at equilibrium is connected to another cell under 
the steady heat conduction is discussed. This solution apparently shows that steady-state thermodynamics 
proposed by Sasa and Tasaki cannot be used in a naive setup. The preliminary result of 
our simulation based on molecular-dynamics for nonequilibrium Knudsen effect is also presented to verify the theoretical
argument.

\end{abstract}

\section{Introduction}

Kinetic theory of dilute gases has played crucial roles in the history
 of nonequilibrium statistical mechanics.
The theoretical treatment of linear transport problems is well established 
 including the derivation of  Navier-Stokes equation and the relation 
between Boltzmann equation and the linear-response
theory through a number of studies of 
Boltzmann equation\cite{chapman,resibois,cercignani1}.

The transport problems
 under the nonlinear-nonequilibrium circumstance, however,
 is still controversial. In fact, we know Burnett's classical 
work in which he obtained an approximate hydrodynamic equation at 
 the second order nonequilibrium condition\cite{burnett}, 
but we realize
that hydrodynamic equation he obtained
 is unstable for perturbations\cite{bobylev,struchtrup}.  
Therefore, one may suggest
that the validity of Chpaman-Enskog method is so limited that we should 
use Grad expansion\cite{grad} 
to discuss the higher order perturbations such as 
Burnett order and super-Burnett order\cite{struchtrup}. 
On the other hand, when we obtain the steady solution of Burnett
equation, the stability of the steady inhomogeneous solution 
may be different from
that of dynamical Burnett equation for homogeneous perturbation.  
Therefore we cannot judge whether
 the calculation at Burnett order based on Chapman-Enskog method is
 useless
 even when Burnett's  hydrodynamic equation is unstable. 
 Unfortunately, these difficulties 
might be regarded as unimportant, because the solution at Burnett order
is practically useless except for few cases such as 
the analysis of shock waves.

In these days, however, the solution of Boltzmann equation 
at Burnett order has attracted much 
interest among researchers who try to extend the linear nonequilibrium 
statistical mechanics to nonlinear nonequilibrium statistical mechanics\cite{jou,sst}.
In fact, since
all of attempts are formulated to match them with the linear
 response theory, their consistency with nonlinear theory like Burnett
 solution should be important.

Recently, 
Kim and Hayakawa\cite{kim03a} have extended Burnett's classical work on
Boltzmann equation and obtained the explicit steady solution of
Boltzmann equation under the steady heat conduction at second (Burnett)
order.  
The accuracy and numerical stability of this solution is confirmed by
Fushiki\cite{fusiki} from his molecular-dynamics simulation.
This stable behavior does not contradict with blown-up tendency of
Burnett equation for perturbation of short wavelength or high frequency, 
because the simulation does not involve any  perturbation with
 short wavelength or high frequency.
Their solution is useful\cite{kim03b} to examine the validity of nonequilibrium
theories such as information theory\cite{jou} and 
the steady-state thermodynamics (SST)\cite{sst}
 based on thermodynamic arguments.  They also succeed to apply the
 solution to calculate the rate of chemical reaction under the
 nonequilibrium circumstance in which the linear nonequilibrium effects
 are canceled from the consideration of symmetries\cite{kim03c}.

The purpose of this paper is to (i) summarize the achievement of Kim and 
Hayakawa, (ii) to discuss the stability
of the steady solution at Burnett order, and (iii) to
introduce the result of the molecular-dynamics (MD) simulation for
nonequilibrium Knudsen effect to examine SST.

The organization of this paper is as follows. In the next section, we
briefly summarize the outline of the solution obtained by Kim and 
Hayakawa\cite{kim03a}.  In section III, we will discuss the stability of 
the solution. In section IV, we will discuss Knudsen effect under a
nonequilibrium condition based on MD simulation.

\section{The outline of the solution by Kim and Hayakawa}

In this section, we briefly summarize the solution of steady Boltzmann
equation under the heat conduction obtained by Kim and
Hayakawa\cite{kim03a}. 
Suppose a system of dilute gas in a steady state whose velocity distribution
function is $f_{\mathrm{1}}=f({\bf r},{\bf v}_{\mathrm{1}})$. 
The steady state Boltzmann equation is written as 
\begin{equation}  
{\bf v}_{\mathrm{1}}\cdot\nabla f_{\mathrm{1}}=
J(f_{\mathrm{1}},f_{\mathrm{2}}),\label{be1}
\end{equation}
where the collision integral $J(f_{\mathrm{1}},f_{\mathrm{2}})$ for hard core molecules is expressed as 
\begin{equation}  
J(f_{\mathrm{1}},f_{\mathrm{2}})=
\int \int \int (f^{\prime}_{\mathrm{1}}f^{\prime}_{\mathrm{2}}-f_{\mathrm{1}}f_{\mathrm{2}})g b db d\epsilon d{\bf v}_{\mathrm{2}}.\label{be2}
\end{equation}
Here the velocity distribution $f_{\mathrm{1}}$ and $f_{\mathrm{2}}$
change to $f^{\prime}_{\mathrm{1}}$ and $f^{\prime}_{\mathrm{2}}$ by a
binary collision. 
The relative velocity of the two molecules before and after the
interaction has the same magnitude $g=|{\bf v}_{\mathrm{1}}-{\bf
v}_{\mathrm{2}}|$.
The relative position of the two molecules is represented by the impact
parameter  $b$ and $\epsilon$ which represents the angle of the plane to which $g$ and
$g^{\prime}$.    
The impact parameter $b$ is identical to explicitly determined by
specifying the kind of molecules.

The velocity distribution function $f_{\mathrm{1}}$ can be expanded as:
\begin{equation}  
f_{\mathrm{1}}=f^{(0)}_{\mathrm{1}}+f^{(1)}_{\mathrm{1}}+f^{(2)}_{\mathrm{1}}+\cdots=f^{(0)}_{\mathrm{1}}(1+\phi^{(1)}_{\mathrm{1}}+\phi^{(2)}_{\mathrm{1}}+\cdots), \label{be3}
\end{equation} 
where the small parameter of the perturbation is Knudsen number
, i.e. mean free path of the molecules  is much less than
the characteristic length of the change of macroscopic
variables. 
$f^{(0)}_{\mathrm{1}}$ in eq.(\ref{be3}) is the local Maxwellian distribution function written as 
\begin{equation}  
f^{(0)}_{\mathrm{1}}=n({\bf r})\left(\frac{m}{2\pi  T({\bf r})}\right)^{\frac{3}{2}}\exp[-\frac{m{\bf v}_{\mathrm{1}}^{2}}{2 T({\bf r})}],\label{be3.5} 
\end{equation} 
where $m$ is mass of the molecules. 
$n({\bf r})$ and $T({\bf r})$ denote the density and the temperature at position ${\bf r}$, respectively. 
Substituting eq.(\ref{be3}) into the steady state Boltzmann equation (\ref{be1}), we
arrive at the following set of equations:  
\begin{eqnarray}  
L[f^{(0)}_{\mathrm{1}}]\phi^{(1)}_{\mathrm{1}}={\bf v}_{\mathrm{1}}\cdot\nabla f^{(0)}_{\mathrm{1}}, \label{be4}
\end{eqnarray}
for $O(K)$ and
\begin{eqnarray}  
L[f^{(0)}_{\mathrm{1}}]\phi^{(2)}_{\mathrm{1}}={\bf v}_{\mathrm{1}}\cdot\nabla f^{(1)}_{\mathrm{1}}-J(f^{(1)}_{\mathrm{1}}f^{(1)}_{\mathrm{2}}),\label{be5}
\end{eqnarray}
for $O(K^2)$, where the linear integral operator $L[f^{(0)}_{\mathrm{1}}]$ is defined as\begin{eqnarray}  
L[f^{(0)}_{\mathrm{1}}]X_{\mathrm{1}}\equiv\int \int \int f^{(0)}_{\mathrm{1}}f_{\mathrm{2}}^{(0)}(X_{\mathrm{1}}^{\prime}-X_{\mathrm{1}}+X^{\prime}_{\mathrm{2}}-X_{\mathrm{2}})g b db d\epsilon d{\bf v}_{\mathrm{2}}. \label{be6}
\end{eqnarray}

It is important to consider the solubility conditions of these integral
equations, for the steady state Boltzmann equation, which are given by
\begin{eqnarray}  
\int \Phi_{i} {\bf v}_{\mathrm{1}}\cdot\nabla f^{(0)}_{\mathrm{1}} d{\bf v}_{\mathrm{1}}=0,\label{be7}
\end{eqnarray}
for $O(K)$, where $\Phi_{i}$ is one of the collision invariants:
\begin{eqnarray}  
\Phi_{1}=1, \quad \Phi_{2}=m{\bf v}_{\mathrm{1}}, \quad \Phi_{3}=\frac{1}{2}m{\bf v}_{\mathrm{1}}^{2}.\label{be8}
\end{eqnarray}
Substituting eq.(\ref{be3.5}) into the solubility condition (\ref{be7}),
it can be shown that $n T$ is uniform. 
We use this result in our calculation. 
The solubility condition for $O(K^2)$ is given by
\begin{eqnarray}  
\int \Phi_{i} {\bf v}_{\mathrm{1}}\cdot\nabla f^{(1)}_{\mathrm{1}} d{\bf v}_{\mathrm{1}}=0.
\label{be9}
\end{eqnarray}

To make the solutions of the integral equations
(\ref{be5}) and (\ref{be6}) definite, five further conditions must be
specified; we identify the density:   
\begin{equation}  
n({\bf r})=\int f_{\mathrm{1}} d{\bf v}_{\mathrm{1}}=\int f^{(0)}_{\mathrm{1}} d{\bf v}_{\mathrm{1}},\label{be10}
\end{equation} 
the temperature:  
\begin{equation}  
\frac{3n({\bf r}) T({\bf r})}{2}=
\int \frac{m{\bf v}_{\mathrm{1}}^{2}}{2}f_{\mathrm{1}} d{\bf v}_{\mathrm{1}}=\int \frac{m{\bf v}_{\mathrm{1}}^{2}}{2}f^{(0)}_{\mathrm{1}} d{\bf v}_{\mathrm{1}}, \label{be11}
\end{equation}
and the mean flow: 
\begin{equation}  
n({\bf r}){\bf u}({\bf r})
=\int {\bf v}_{\mathrm{1}}f_{\mathrm{1}} d{\bf v}_{\mathrm{1}}=\int  {\bf v}_{\mathrm{1}}f^{(0)}_{\mathrm{1}} d{\bf v}_{\mathrm{1}}.\label{be12}
\end{equation} 
Here we assume that no mean flow ${\bf u}=0$ exists in
the system.    

Skipping the derivation of the solution at Burnett order, let us write
the explicit form of the steady solution for hard core molecules as
under the temperature gradient:
\begin{eqnarray}  
f=f^{(0)}[1&-&\frac{4q_{x}}{5 b_{11} n T}(\frac{m}{2 T})^{\frac{1}{2}}\sum_{r\ge 1}r! b_{1r}c_{x}\Gamma(r+\frac{5}{2}) S^{r}_{\frac{3}{2}}({\bf c}^{2})\nonumber\\
&+&\frac{4096m q_{x}^{2}}{5625b_{11}^{2}n^{2}T^{3}}\{\sum_{r\ge 2}r! b_{0r}\Gamma(r+\frac{3}{2})S^{r}_{\frac{1}{2}}({\bf c}^{2})\nonumber\\
&+&\sum_{r\ge 0}r! b_{2r}(2c_{x}^{2}-c_{y}^{2}-c_{z}^{2})\Gamma(r+\frac{7}{2})S^{r}_{\frac{5}{2}}({\bf c}^{2})\}
], \label{be46}
\end{eqnarray}
where $b_{1r}$, $b_{0r}$ and $b_{2r}$ for $r\le 7$ can be determined as
listed in
Tables.\ref{b0r}. Note that we have confirmed the
convergence of series $b_{kr}$ by changing $r$. Here
$q_{x}$ represents $x-$component of the heat-flux in the cell on the 
right hand side in
the nonequilibrium steady state which corresponds to the $x$ component of
 the heat flux in eq.(\ref{be39}). 
In eq.(\ref{be46}), $\Gamma(x)$ is the Gamma function, ${\bf c}={\bf
v}\sqrt{m/2T}$,  and $S^p_k(X)$ is the Sonine
polynomial defined by 
\begin{eqnarray}  
(1-\omega)^{-k-1}e^{-\frac{X\omega}{1-\omega}}=\sum_{p=0}^{\infty} \Gamma(p+k+1) S_{k}^{p}(X)\omega^p. 
\label{be13.5}
\end{eqnarray}
This kind of analysis, of course, is possible to two-dimensional dilute 
hard-core gases. Kim\cite{kim04} has already obtained the result to
compare his result with MD simulation.

In the derivation 
the solubility conditions for $O(K^2)$ is important.
Thus, eq.(\ref{be9}) leads to
the condition: 
\begin{eqnarray}  
\nabla \cdot {\bf q}^{(1)}=0, \label{be38}
\end{eqnarray}
where ${\bf q}^{(1)}$, i.e. the heat flux for $f^{(1)}_{\mathrm{1}}$ can be obtained as  
\begin{equation}  
{\bf q}^{(1)}=\int^{\infty}_{-\infty}d{\bf v}_{\mathrm{1}}\frac{m{\bf v}_{\mathrm{1}}^{2}}{2}{\bf v}_{\mathrm{1}} f^{(1)}_{\mathrm{1}}
=-b_{11}\frac{75}{64\sigma^2}\left(\frac{ T}{\pi m}\right)^{\frac{1}{2}}
\nabla T, \label{be39}
\end{equation}  
with $b_{11}$ listed in Table.\ref{b0r}. 
It must be emphasized that ${\bf q}^{(2)}$ should satisfy
\begin{equation}\label{q_2}
{\bf q}^{(2)}=0
\end{equation}
 from the
symmetry consideration of $f^{(2)}$.
Thus, the solubility conditions for
$f^{(2)}_{\mathrm{1}}$ of the steady state Boltzmann
equation lead to the fact that the heat flux is
constant to the second order. 
From eqs.(\ref{be38}) and (\ref{be39}), we also obtain an important relation
between $\left(\nabla T\right)^{2}$ and $\nabla^{2} T$ as 
\begin{eqnarray}
\frac{\left(\nabla T\right)^{2}}{2T}+\nabla^{2} T=0.  
\label{be39.5}
\end{eqnarray} 
From eq.(\ref{be39.5}), the term of $\nabla^{2} T$ can be replaced by the terms of $\left(\nabla T\right)^{2}$. 

\begin{figure}[htbp]
\epsfxsize=8cm
\centerline{\epsfbox{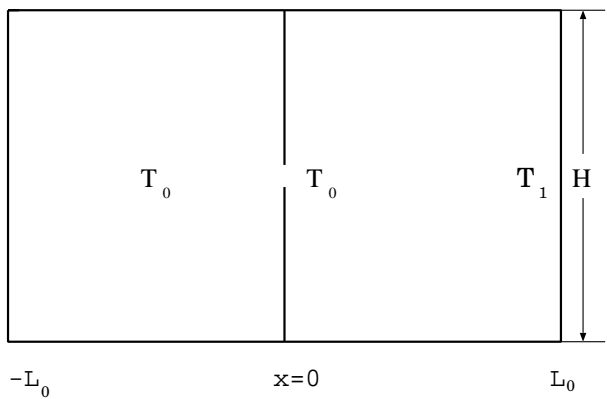}} 
\caption{Setup of nonequilibrium Knudsen effect.
}
\label{sst}
\end{figure}

Amongst several applications of the solution, in this paper, we
 emphasize that the solution can examine the validity of SST proposed by 
 Sasa and Tasaki\cite{sst}. 
They propose a non-trivial nonequilibrium effect in a
 simple setup as follows( Fig. \ref{sst}).
 There are two cells whose size along $x$ axis is $L_{0}$.  
Both of the cells are filled with dilute gases and connected to each other by a small
 hole at $x=0$ whose linear dimension along $x$ axis and its diameter
 are less than the mean free path.
The cell on the left hand side is at equilibrium of
temperature $T_{0}$, while the cell on the right hand side is in a
 nonequilibrium steady state under a temperature gradient caused by the right wall
 at temperature $T_{1}$ at $x=L_{0}$ and the thin mid-wall at temperature
 $T_{0}$.

Sasa and Tasaki predicted the following results\cite{sst}. 
When the system shown in Fig.\ref{sst} is in a steady state, the osmosis
$\Delta P$ 
\begin{eqnarray} 
\Delta P&\equiv&P_{xx}-P_{0}, \label{be151} 
\end{eqnarray}
always positive, where $P_0$ is the pressure in the equilibrium cell
and $P_{xx}$ is the $xx$-component of
 the pressure tensor $P_{ij}$ in the nonequilibrium cell which is defined by
\begin{equation}
P_{ij}=\int d{\bf v }m (v_i-u_i)(v_j-u_j)f({\bf v}).
\end{equation}
In addition, 
there is a relation among $n(0)$, $P_{xx}$, $n_{0}$ and $P_{0}$ as 
\begin{eqnarray}  
\frac{n(0)}{n_{0}}=\left(\frac{\partial P_{xx}}{\partial P_{0}}\right)_{T_{0},J}, \label{be53}
\end{eqnarray}
where $n(0)$ is the density of the cell in the nonequilibrium steady
state around the wall and $n_{0}$ is the density of the
cell at equilibrium. 

A nonequilibrium 
steady state is achieved when the mean mass flux at the hole is zero.
Neglecting the Knudsen layer effect, i.e. the slip effect 
around the 'wall'\cite{cercignani1,landau}, 
we can calculate the balance of mass flux as
\begin{eqnarray}\label{int}  
\int^{\infty}_{0}dv_{x}\int d^{2}v_{\bot}mv_{x}f_{0}+
\int^{0}_{-\infty}dv_{x}\int d^{2}v_{\bot}mv_{x}f|_{x=0}=0, \label{be48}
\end{eqnarray}
where $v_{\bot}$ represents the components of the velocity which are
orthogonal to $v_{x}$, i.e. $v_{y}$ and $v_{z}$. 
From eq.(\ref{be48}), we can obtain 
the relation between the density of the cell in the
nonequilibrium steady state around the hole $n(0)$ and that of the
cell at equilibrium $n_{0}$ as
\begin{eqnarray}  
n(0)=n_{0}[1+\lambda_{n}\frac{mJ_{x}^{2}}{n_{0}^{2}T^{3}}], \label{be50}
\end{eqnarray}
to the second order, where
 $\lambda_{n}\simeq 8.21\times 10^{-2}$ in the approximation until 7th
 order expansion of $b_{kr}$.

Similarly we can calculate
pressure tensor $P_{ij}$  as
\begin{eqnarray}  
P_{ij}&=& n T[\delta_{ij}+\lambda_{P}^{ij}\frac{mJ_{x}^{2}}{n_{0}^{2}T^{3}}],\label{be51}
\end{eqnarray}
with the unit tensor $\delta_{ij}$ and the number tensor
$\lambda_{P}^{xx}=-4.600\times 10^{-2}$ and
$\lambda_{P}^{xy}=2.300\times 10^{-2}$.

From substitution of eqs.(\ref{be50}) and (\ref{be51}) into eq.(\ref{be53})
leads to the relation $\lambda_{P}^{xx}/\lambda_{n}=-2$ in SST.
On the other hand, our results conflict with this
relation: our results become $\lambda_{P}^{xx}/\lambda_{n}=-0.5604$
for  hard core molecules. 
This is a negative result to SST, though the treatment of the wall in
our calculation is too primitive. We also indicate that our treatment in 
eq.(\ref{be50}) assumes that we can use the local distribution
function because of small Knudsen number, 
though, for dilute gases with finite temperature gradient, 
the place with the mean-free path away from the hole has different
temperature from that at the hole because Knudsen number should be finite.

\section{Stability of the solution}

In this section, let us discuss the stability of the second order
solution in eq.(\ref{be46}). Honestly speaking, it is difficult to
obtain the complete stability analysis of the problem. Therefore, here,
we focus on the stability of five hydrodynamic modes which are
zero-eigen modes of linearized Boltzmann operator and most
dangerous ones in the stability.

We should recall that the steady solution with the temperature field
$T(x)$ is obtained under the
assumption of  the zero mean (hydrodynamic)
flow. With the aid of eq.(\ref{q_2}) one of the hydrodynamic equations
corresponding to the solution is the steady diffusion equation:
\begin{equation}\label{diffusion}
\nabla\cdot(\kappa \nabla T(x))=0, \qquad \kappa=b_{11}\frac{75}{64\sigma^2}
\left(\frac{ T}{\pi m}\right)^{1/2}
\end{equation}
where $\kappa$ is the heat conductivity. Another hydrodynamic
equation is nothing but equilibrium condition of hydrostatic pressure: 
\begin{equation}
\nabla(n_0 T(x))=0
\end{equation}
where $n_0({\bf r})=\int d{\bf v} f$ 
is the density determined from the solution.

When we keep the constraint of the static solubility conditions
(\ref{be38}) and (\ref{q_2}), 
there is no room to appear ${\bf q}^{(2)}$ in the
hydrodynamic equations. Even when the perturbation violates the
static solubility condition, 
the possible second order heat current is only
\begin{equation}\label{q_2new}
{\bf q}^{(2)}
= \alpha
\frac{\mu^2}{n_0m T}u_x\frac{d}{dx} T(x)
\end{equation}
in the perturbation,
where $\alpha=45/8$, and $\mu$ is the shear viscosity.
It should be noted that the above expression is  from 
 eq.(15.3.6) in Chapman-Cowling\cite{chapman} with neglecting the
nonlinear terms of the velocity field. 
We also note that the perturbed pressure tensor has the form of
Navier-Stokes order, because the velocity field itself is a perturbed
quantity. Thus, the linearized
hydrodynamics has a similar form as that of Navier-Stokes equation 
except for  a term of the heat current in eq.(\ref{q_2new}).

Nevertheless, it is
 still not easy to discuss the complete stability of hydrodynamic
modes around the steady Burnett solution.
Let us discuss the perturbation of density and temperature
\begin{equation}\label{hydro-vari}
n=n_0(1+\rho({\bf r},t)),  \quad \Theta=T(x)(1+\theta({\bf r},t)),
\end{equation}
where $\Theta$ is the temperature after we include the perturbation.
Here we keep that the macroscopic velocity does not exist for the sake
of simplificity.
In this case it is easy to confirm that the linearized hydrodynamic
equations are reduced to $\partial_t\rho=0$ and
\begin{equation}
\frac{3}{2}n_0\partial_t\theta=\kappa^*T^{3/2}\nabla^2\theta-
\frac{5}{2}q_x\partial_x\theta-\frac{q_x}{2T}(1+\frac{q_x}{\kappa^*\sqrt{T}})\theta
\end{equation}
where the heat conductivity is assumed to be $\kappa=\kappa^*\sqrt{T}$
with $\kappa^*=75b_{11}/(64\sigma^2\sqrt{\pi m})$.
This equation apparently shows that the solution of $\theta$ is relaxed
to zero for all wavelength, as time goes on. 
Thus, the steady solution of Boltzmann
equation at Burnett order is stable for the perturbation
(\ref{hydro-vari}) without macroscopic flow.

\section{Molecular-Dynamics Simulation}

In this section, we show our preliminary result of two-dimensional
molecular-dynamics (MD) simulation to examine the Knudsen effect under
the heat conduction.  Although we adopt two-dimensional systems, MD can
check the validity of their 
implicit assumptions of the calculation by Kim and
Hayakawa\cite{kim03a} where their calculation neglect the effects of 
the thermal wall at $x=0$ and use the distribution function at $x=0$ in
eq. (\ref{int}).

We prepare two situations where $T_1/T_0=4$ and $T_1/T_0=1/4$. Number of 
total particles is 10,000 for both cases. Neglecting the non-Gaussian
correction  in velocity distribution function, the mean-free paths
$l=1/(2\sqrt{2}n\sigma)$ for the center regions of 
four cells are respectively evaluated as 
$l_{neq}=11.24d$ and $l_{eq}=17.06d$ for the case $T_1/T_0=1/4$ and
$l_{neq}=21.40d $ and $l_{eq}=9.534d$ for $T_1/T_0=4$, where $l_{eq}$
and $l_{neq}$ represent the mean-free paths for an equilibrium cell and
a nonequilbrium cell, respectively. 
Figure 2 shows the result of both situations. $L_0\simeq1071 d$ and
the height $H\simeq 268.6 d$. The linear dimension along x-axis
 of the hole is 0.40845
$d$ and the diameter of the hole is $1.113 d$. This system has the
area fraction $\phi=0.02730$.
The vertical axis in Fig.2 plots the pressure divided by the
density and temperature of equilibrium cell $T_0$. The horizontal axis
shows the time of simulation measured by the number of collisions for
each particle. We adopt the
diffusive boundary condition for the walls at $x=0$ and $x=\pm L_0$, and 
the periodic boundary condition at $y=\pm H/2$.

\begin{figure}[htbp]
\epsfxsize=12cm
\centerline{\epsfbox{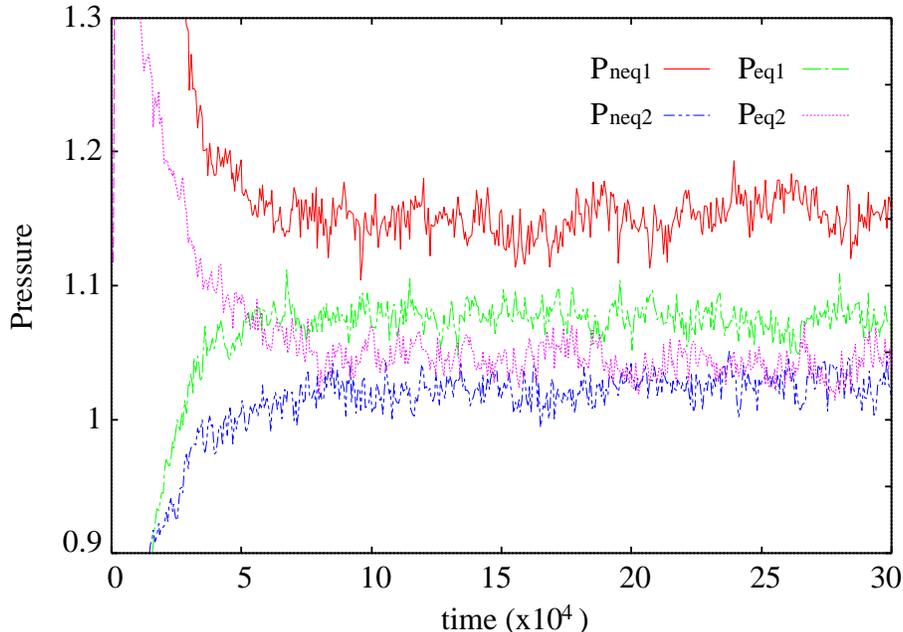}} 
\caption{
The time evolution of pressures in MD simulation where $P_{neq1}$ and
 $P_{eq1}$ are the result of the simulation for $T_1/T_0=4$, and
 $P_{neq2}$ and $P_{eq2}$ are the result for $T_1/T_0=1/4$.   
}
\label{nishino}
\end{figure}

In Fig.2,
the suffices 1 and 2 in the figure correspond to the cases of 
$T_1/T_0=4$  and $T_1/T_0=1/4$. 
Our result is contrast to those of  both SST\cite{sst} and the calculation by 
Kim and Hayakawa\cite{kim03a}.
Here, the osmosis $\Delta P$ depends on the gradient of
temperature in which $\Delta P$ becomes positive for $T_1/T_0=4$ and
$\Delta P$ becomes negative for $T_1/T_0=1/4$.
This simulation suggests that both SST and the calculation by 
Kim and Hayakawa cannot be used for Knudsen effect in physical situations.

There are two possibilities to have linear dependence of the osmosis
$\Delta P$ on the
temperature gradient: (i) Knudsen's layer effect is important on the
thermal wall, and (ii) the assumption of small Knudsen number is
violated. As a result of finiteness of Knudsen number which is
around 1/50-1/100, 
the temperature at the place deviated from the
hole by the mean-free path is significantly different from the
temperature at the hole. We believe that the second possibility plays
a dominant role, because the gap of the pressure is enhanced when the 
nonequilibrium cell has higher temperature, {\it i.e. } larger mean-free 
path $l_{neq}=21.40d$ than that for low temperature nonequilibrium cell
$l_{neq}=11.24d$.  The details of MD simulation
will be reported elsewhere.

\section{Conclusion}

In conclusion, we have explained our recent finding of a new explicit
solution of Boltzmann equation under the steady heat conduction. We
apply the solution to examine the validity of SST and information
theory, and find that both of them are not appropriate as they are.
We also discuss the stability of the steady solution we have
obtained, and confirmed stability at least for the case of no systematic 
flow. Through MD simulation of Knudsen effect under the heat
conduction, we have obtained the result that $\Delta P$ depends on the
direction of the temperature gradient. This indicates that our naive
application of the steady solution of Boltzmann equation to Knudsen
effect is not accurate. This also suggests that the basic assumption of
SST cannot be valid if the influence of physical wall exists.

\vskip 0.5cm
Acknowledgment:
The authors wish to thank M. Fushiki for fruitful discussion and 
for his sending us the unpublished result of his simulation. 
This work is partially supported by the Grant-in-Aid of Ministry of
Education, Culture, Sports, Science and Technology(MEXT), Japan (Grant
No. 15540393) and the Grant-in-Aid for the 21st century COE 'Center for
Diversity and Universality in Physics' from MEXT, Japan.

\begin{table}[htbp]
\epsfxsize=8cm
\caption{\label{b0r}The number constants $b_{0r}$ in
 eq.(\ref{be46})}
\begin{tabular}{|c|c|c|c|}   
{$r$}&{$b_{0r}$}&{$b_{1r}$}&{$b_{2r}$} \\ \hline
{$0$}&{-}&{-}&{$-3.320\times 10^{-2}$}
 \\ \hline
{$1$}&{-}&{$1.025$}&{$-1.276\times 10^{-1}$}
 \\ \hline
{$2$}&{$4.380\times 10^{-1}$}&{$4.892\times 10^{-3}$}&{$6.414\times 10^{-2}$}
 \\ \hline
{$3$}&{$-5.429\times 10^{-2}$}&{$3.715\times 10^{-3}$}&{$5.521\times 10^{-3}$} \\ \hline
{$4$}&{$-4.098\times 10^{-3}$}&{$2.922\times 10^{-4}$}&{$4.214\times 10^{-3}$}
 \\ \hline
{$5$}&{$-3.184\times 10^{-4}$}&{$2.187\times 10^{-5}$}&{3.106$\times 10^{-5}$} \\ \hline
{$6$}&{$-2.087\times 10^{-5}$}&{$1.492\times 10^{-6}$}&{$1.861\times
 10^{-6}$}
 \\ \hline
{$7$}&{-}&{$8.322\times 10^{-8}$}&{-} \\
\end{tabular}
\end{table}


\begin{references}
\bibitem{chapman} S. Chapman and T. G. Cowling, The Mathematical Theory
 of Non-Uniform Gases , Third Edition (Cambridge University Press, 1970).
\bibitem{resibois} P. R\'esibois and M. de Leener, Classical Kinetic
 Theory of Fluids (John Wiley $\&$ Sons, New York, 1977).
\bibitem{cercignani1} C. Cercignani, \textit{Mathematical Methods in
	Kinetic Theory}(Plenum Press, New York, 1990).
\bibitem{burnett} D. Burnett, {\it The Distribution of Molecular
 Velocities and the Mean Motion in a Non-Uniform Gas},
 Proc. Lond. Math. Soc. 1935, {\bf 40}
	(Nov.15), 382-435.
\bibitem{bobylev} A. V. Bobylev, {\it The Chapman-Enskog and Grad Method 
 for solving Boltzmann equation}, Sov. Phys. Dokl. 1982, {\bf 27}, 29.
\bibitem{struchtrup} H. Struchtrup and M. Torrihon, {\it Regularization
 of Grad's 13 moment equation: Derivation and linear analysis},
 Phys. Fluids, 2003, {\bf 15} (9), 2668-2680.
\bibitem{grad} H. Grad, {\it On the kinetic theory of rarefied gas},
 Commun. Pure Appl. Math. 1949, {\bf 2} 331-407.
\bibitem{jou} J. Casas-V\'arquez and D. Jou, {\it Temperature in
 non-equilibrium states: a review of open problems and current proporsal
 }, Rep. Prog. Phys. 2003, {\bf 66}, 1937-2023.
\bibitem{sst} S. Sasa and H. Tasaki, {\it Steady state thermodynamics}, 
cond-mat/0411052, 2004. See also their previous version in cond-mat/0108365.
\bibitem{kim03a} Kim Hyeon-Deuk  and H. Hayakawa, {\it Kinetic Theory of
 a Dilute Gas System under Steady Heat Conduction}
J. Phys. Soc. Jpn. 2003 
 , {\bf 72} (8), 1904-1916.
\bibitem{fusiki} M. Fushiki, private communications. 
(and his unpublished result). 
\bibitem{kim03b} Kim Hyeon-Deuk  and H. Hayakawa,
{\it Test of information theory on the Boltzmann equation}
J. Phys. Soc. Jpn. 2003 {\bf 72}, (10), 2473-2476.
\bibitem{kim03c} Kim Hyeon-Deuk  and H. Hayakawa, {\it Contributions of
 steady heat conduction to the rate of chemical reaction},
Chem. Phys. Lett. 2003, {\bf 372}, (Nos.3-4) 314-319.
\bibitem{kim04} Kim Hyeon-Deuk, in preparation.

\bibitem{landau} E. M. Lifshitz and Pitaevski, \textit{Physical
	Kinetics}(Pergamon, Oxford, 1981).

\end{references}
\end{document}